\documentclass[journal]{IEEEtran}
\usepackage{cite}
\usepackage{color}
\usepackage{amsmath,amssymb,amsfonts}
\usepackage{algorithm}
\usepackage{algorithmic}
\usepackage{graphicx}
\usepackage{textcomp}
\usepackage{subfigure}
\usepackage{multirow}
\def\BibTeX{{\rm B\kern-.05em{\sc i\kern-.025em b}\kern-.08em
    T\kern-.1667em\lower.7ex\hbox{E}\kern-.125emX}}

\newtheorem{remark}{\bf Remark}
\newtheorem{definition}{\bf Definition}[section]
\newtheorem{assumption}{\bf Assumption}[section]
\newtheorem{lemma}{\bf Lemma}[section]

\newtheorem{theorem}{\bf Theorem}[section]
\newtheorem{corollary}{\bf Corollary}[section]

\newcommand{\Real}{\mathbb R}
\newcommand{\Tran}{\mathrm T}
\newcommand{\norm}[1]{\left\Vert#1\right\Vert}

\begin{document} 
\title{Suboptimal Safety-Critical Control for Continuous Systems Using Prediction-Correction Online Optimization}
\author{Shengbo Wang, Shiping Wen,~\IEEEmembership{Senior Member,~IEEE}, Yin Yang, Yuting Cao, Kaibo Shi, \\ and Tingwen Huang,~\IEEEmembership{Fellow,~IEEE}
\thanks{This publication was made possible by NPRP grant: NPRP 9-466-1-103 from Qatar National Research Fund. The statements made herein are solely the responsibility of the authors. \textit{(Corresponding authors: Shiping Wen.)}
}
\thanks{S. Wang is with the School of Computer Science and Engineering, University of Electronic Science and Technology of China, Chengdu 611731, China (e-mail: shnbo.wang@foxmail.com).}
\thanks{S. Wen is with Australian AI Institute, Faculty of Engineering and Information Technology, University of Technology Sydney, NSW 2007, Australia (shiping.wen@uts.edu.au).} 
\thanks{Y. Yang and Y. Cao are with College of Science and Engineering, Hamad Bin Khalifa University, 5855, Doha, Qatar (email: ycao@hbku.edu.qa, yyang@hbku.edu.qa).}
\thanks{K. Shi is with School of Information Science and Engineering, Chengdu University, Chengdu, 611040, China, (email: skbs111@163.com).}
\thanks{T. Huang is with Science Program, Texas A \& M University at Qatar, Doha 23874, Qatar (e-mail: tingwen.huang@qatar.tamu.edu).}
}

\maketitle

\begin{abstract}
This paper presents an innovative efficient safety-critical control scheme for nonlinear systems by combining techniques of control barrier function (CBF) and online time-varying optimization. The idea lies in that when directly calling the complete optimization solvers used in the CBF method, such as CBF-based quadratic programming (CBF-QP), is computationally inefficient for complex tasks, the suboptimal solutions obtained from online learning techniques can be an alternatively reliable choice for ensuring both efficiency and control performance. By using the barrier-based interior point method, the original optimization problem with CBF constraints is reduced to an unconstrained one with approximate optimality. Then, Newton- and gradient-based continuous dynamics are introduced to generate alternative cheap solutions while ensuring safety. By further considering the lag effect of online tracking, a prediction term is added to the dynamics. In this way, the online cheap solutions are proven to exponentially converge to the time-varying suboptimal solutions of the interior point method. Furthermore, the safety criteria are established, and the robustness of the designed algorithms is analyzed theoretically. Finally, the effectiveness is illustrated by conducting two experiments on obstacle avoidance and anti-swing tasks.
\end{abstract}


\section{Introduction}
\label{sec:introduction}
\IEEEPARstart{C}{ontrol} of autonomous systems attracts more interest from researchers in recent years. The study on optimization of the control performance, such as linear quadratic regulator (LQR) optimal control and model predictive control (MPC), has been successfully developed and practically applied in industry \cite{MPC_constraints,explicit_LQR_constrained,RL_adp_saturation,system_level_sythesis,SHI2021170,tsmcs_MPC_state_constraint,shnbo_PETC_optimalcontrol,DingWang}. When integrating constraints in the control process, designing and optimizing a valid controller become harder, especially in online control scenarios. As simple cases, the state constraints and input saturation constraints are studied in the literature \cite{explicit_LQR_constrained,RL_adp_saturation, system_level_sythesis,MPC_constraints,tsmcs_MPC_state_constraint}, where the constraints are commonly assumed fixed. This assumption does not apply to cases with safety constraints that are naturally varying, and dependent on the system states and external inputs. To cope with this, long-term optimization methods such as MPC may suffer a heavy computational burden. Instead, some myopic optimization methods can better balance the constraint satisfaction and computation complexity for safe control tasks.

\par A key challenge in safe control tasks is to quantity the safety measure in an easy-to-handle way \cite{safeandfast_tsmcs}. The control barrier function (CBF) method quantifies safety as CBF-based constraints which are linear in control input. The idea comes from the invariant set, inside which any state starting will keep inside it forever \cite{nonlinear_systems}. This is a promising way to effectively keep systems safe during the control process \cite{cbfdefine}. CBF method is tractable since the constraints are linear and the objective functions can be manually designed convex, such as quadratic programming (QP). As a myopic method, CBF method only computes the safe control input for each step, resulting in quite an efficient scheme for online implementation. In applications, CBF method has been used for obstacle avoidance, bipedal walking, and robotic grasping \cite{CBF_dynamic_robotic_TAC,bipedal_walking,CBF_robot_grasping}. Developing CBF for uncertain disturbances and systems with a higher degree of freedom has been investigated in \cite{robust_CBF,exponential_CBF,high_order_CBF}. 
By introducing the control Lyapunov function (CLF), a synthetic CBF-CLF method is presented as an effective safe autonomous control framework \cite{cbfdefine}. As model uncertainty will badly affect the performance of the CBF-CLF method, safe learning, adaptive control, and the reduction of conservativeness towards the existence of uncertainty are further studied in \cite{CBF_for_bayesian,RaCBF_IEEECSL,WangACBF_FT}.

\par The CBF method needs to solve a convex constrained problem at each control step. Therefore, for continuous systems, the selected solver should be highly efficient to generate almost continuous-time control input without breaking the safety constraints \cite{self_trigger_CBF}. The interest in the computational cost of the CBF method is twofold in this paper. On the one hand, the CBF-QP cannot be integrated with warm-started techniques, such as the active set method \cite{active_set_MPC,active_set_QP_robot}, due to the varying property of safety constraints. Therefore, the computation must start over at every step. However, since most safe solutions are locally continuous, the computational efficiency can be improved by exploiting the historical information of the former steps. On the other hand, CBF method does not consider the evolution of system dynamics, bringing inevitable lag or delay in control effect. The CBF method always provides suboptimal solutions due to varying systems, varying constraints, and even varying objectives. Since the system models have been utilized in the CBF method, more exploitation for predicting the evolution can be useful to compensate for the delayed and suboptimal effect. CBF-QP can be viewed as an instance of the time-varying constrained online convex optimization problems \cite{time_varying_CO_IEEE,tingwen_TVO_constraint}, where the gradient flow, prediction, and feedback bandit have been introduced for better solving these problems in the literature \cite{prediction_correction_unconstrained,OCO_bandit_feedback}. A slightly strong requirement of CBF-QP is, the solutions should always stay in the admissible region, such that the safety constraints will not be violated during the optimization process. As is well known, the interior point method seeks to optimize the current solutions from the interior area of the feasible set, making it possible to solve the CBF-QP \cite{convex_book}. In \cite{prediction_correction_interior_point}, a modified interior point method is proposed to track the continuous solutions of the time-varying optimization problem, where the exponential convergence result is obtained by additionally considering the Newton-based prediction term. Since the Newton method needs to compute the Hessian and its inverse, in \cite{prediction_correction_constrained_projection}, a Hessian-free method is proposed by introducing a projection operation. It is noted that this method only applies to cases where projection operation is more efficient than the computation of inverse. In this paper, the ideas of the interior point method and online convex optimization are leveraged to solve the CBF-QP problem for better efficiency, therefore potentially scalable to high-dimensional systems.

\par From the above observations, in this paper, a safe controller is designed by the CBF method with improved efficiency using time-varying convex optimization techniques. The same idea is followed to deal with the CLF-based stabilizing control problem \cite{tracking_fixedtime_TVOF}. Differently, safety is more difficult to be guaranteed compared to stability considered in \cite{tracking_fixedtime_TVOF}. The main contributions are listed as follows.
\begin{enumerate}
    \item The interior point method is adopted to generate the suboptimal solutions for the CBF method continuously. By using the prediction-correction method \cite{prediction_correction_interior_point}, two descent-based algorithms are given with theoretical support that both can exponentially converge to the suboptimal solutions of the original CBF method. 
    \item The safety criteria for continuous systems steered by the generated solutions from the two algorithms are established. In addition, the robustness against unknown bounded disturbances is analyzed. Safety and input-to-state tracking stability of the two algorithms towards disturbances are theoretically ensured.
    \item The online implementation of the algorithms is presented. Experiments on obstacle avoidance and anti-swing control are conducted to illustrate the effectiveness. 
    The efficiency of the proposed algorithms is highlighted in comparison to the original CBF method.
\end{enumerate}

\par \emph{Notations.} For vector $x$ and matrix $A$, $\norm{x}$ and $\norm{A}$ are the Euclidean norm and induced matrix norm respectively. For vector $x\in \Real^n$ and a differentiable scalar function $h(x)$, the Lie derivative of a vector field $f(x):\Real^n \to \Real^n$ is represented by $L_h f(x) = \frac{\partial h^{\Tran}}{\partial x} f(x)$. The $n$-dimensional identity matrix is denoted by $\mathbf{I}_n$, and the $m$-dimensional vector with all elements zero is $\vec{0}_m$. The gradient of function $f(x,y)$ with respect to $x$ is denoted by $\nabla_x f(x,y)$, while the partial gradient of $\nabla_x f(x,y)$ with respect to $y$ is denoted by $\nabla_{xy}f(x,y)$.

\section{Problem Formulation}
\label{section_problem_formulation}
The nonlinear control affine systems considered in this paper take the form as follows:
\begin{equation}
    \dot x(t) = f(x(t)) + g(x(t)) u,\quad t \ge t_0 \label{system_dynamic}
\end{equation}
where $x(t)\in\Real^{n}$ is the system state, $u\in \Real^m$ is the control input, nonlinear functions $f:\Real^n \to \Real^n$ and $g:\Real^n\to \Real^{n\times m}$ represent the locally Lipschitz continuous drift and input dynamics respectively. In the safety-critical environment, define the safe region of both state and control input as $\mathcal{X}\subset \Real^{n}$ and $\mathcal{U}\subset\Real^m$. The safety requirement is to ensure $x(t)\in \mathcal{X}$ and $u\in \mathcal{U}$ throughout starting from $t_0$. The controller must be carefully designed to achieve the control objective without breaking the safety constraints. For the sake of brevity, systems \eqref{system_dynamic} should be of one degree of freedom (DOF).
Concretely, for the state $x^\prime$ limited in a safe region, $\dot x^\prime$ should be linear to $u$. Note that $x^\prime$ can be a sub-vector of $x$. The detailed examples and extensions can be found in \cite{exponential_CBF}. As a slight abuse, $x$ is used as a short version of $x(t)$ in the following content.

\par This paper focuses on presenting an easy-to-implement algorithm to efficiently design the controller under safety-critical environments. 
In what follows, the CBF-CLF method and an efficient online convex optimization method are reviewed.

\subsection{CBF-CLF method}
A CBF $h(x)$ can give a sufficient condition for a pre-defined set to be forward invariant, known as the safe set. The specific definition of $h(x)$ is given below.

\begin{definition}[CBF and Safe Set \cite{cbfdefine}]
	\label{definition_CBF}
	For a closed convex set $\mathcal{C}$ and systems \eqref{system_dynamic}, a zeroing CBF is defined by a continuously differentiable function $h(x):\Real^n \to \Real$. Let $\mathcal{C}\triangleq \left\{ x \vert h(x)>0 \right\}$, then a valid CBF should satisfy that $\forall x\in \mathcal{C}$,  $\alpha>0$ and
	\begin{equation}
        \sup_{u\in \mathcal{U}} \left[\frac{\partial h}{\partial x}(x) \left( f(x) + g(x) u \right)\right] + \alpha h(x) \ge 0. \label{eqref_definitionCBF}  
	\end{equation} 
	If there exists any local Lipschitz continuous controller rendering \eqref{eqref_definitionCBF}, then $\mathcal{C}$ is forward invariant and systems \eqref{system_dynamic} are safe in $\mathcal{X} \subseteq \mathcal{C}$.
\end{definition}

\par In the original definition of CBF, $\alpha$ in \eqref{eqref_definitionCBF} denotes the extended class $\mathcal{K}$ function. It is noted that $h(x)$ is simplified to be a class $\mathcal{K}$ function considered in this paper. Similarly, the CLF constraint is derived to manipulate a valid Lyapunov function $V$ decreasing at an exponential rate. Concretely, $\dot V \leq -\beta V$ should be satisfied, where $\beta > 0$ represents the convergence rate. Hence, a Lyapunov function $V \ge 0$ can be constructed such that the control target is achieved when $V = 0$. The explicit expression of a CLF is given by
\begin{equation*}
        \inf_{u\in \mathcal{U}} \left[\frac{\partial V}{\partial x}(x) \left( f(x) + g(x) u \right)\right] + \beta V(x) \leq 0.
\end{equation*}

\par The CBF and CLF constraints are put together in convex constrained problems to ensure both stable control and safety. Since safety is the most critical principle, the CLF condition can be slacked if the two kinds of constraints conflict. In all, the general CBF-CLF programming problem is formulated as
\begin{gather}
    u^*(x(t), t) = \arg\min_{(u,\delta)\in \Real^{m+1}} f_0(u, \delta, \vartheta(x(t), t)) \label{constraint_ploblem_origin}\\
    L_f h(x) + L_g h(x) u  + \alpha h(x) \ge 0,\tag{\ref{constraint_ploblem_origin}{a}} \label{constraint_ploblem_origin_cbf}\\
    s.t. ~ L_f V(x) + L_g V(x) u + \beta V(x) \leq \delta, \tag{\ref{constraint_ploblem_origin}{b}} \label{constraint_ploblem_origin_clf}\\
    u_{min} \leq u \leq u_{\max}\tag{\ref{constraint_ploblem_origin}{c}} \label{constraint_ploblem_origin_saturation},
\end{gather}
where $u^*$ is the optimal time-varying solution, $f_0$ represents the user-defined objective function, $\vartheta(x(t), t)$ implies that $f_0$ can also depend on external input $\vartheta$, and $\delta$ is the slack variable for feasibility when conflict happens. The external input can be state-dependent and time-varying. Furthermore, the control input is bounded by $u_{\min}$ and $u_{\max}$.

\par Without losing generality, let $y = \left[ u, \delta \right] $ and $f_0(y, \vartheta)$. As a typical form of the CBF-CLF methods, one can choose $f_0 = \norm{u - u_{ref}}^2 + \lambda \delta^2$, in which $u_{ref}$ is the nominal control input and $\lambda>0$. Then, problem \eqref{constraint_ploblem_origin} can be reformulated in a compact form as
\begin{gather}
    y^*(x(t)) = \arg\min_{y \in \Real^{m+1}} y^\Tran Q y - 2 H^\Tran(x(t)) y \label{constraint_ploblem_QP_saturation}\\    
    s.t. ~ F(y,x(t)) \leq 0 \tag{\ref{constraint_ploblem_QP_saturation}{a}}\label{constraint_ploblem_QP_constraints},
\end{gather}
where $F(y,x(t)) = A^\Tran(x(t))y - B(x(t))$, and 
\begin{gather*}
    Q =   \left[\begin{matrix} 
    \mathbf{I}_m & {0}\\
    \vec{0}_m & \lambda
    \end{matrix}\right], \quad H(x(t)) = \left[\begin{matrix} 
    -2 u_{ref}(x(t))\\
    0
    \end{matrix}\right], \\
    A(x(t)) =  \left[\begin{matrix}
    L_g V(x) & -L_g h(x)& \mathbf{I}_m & -\mathbf{I}_m \\
    -1 & 0 & \vec{0}_m & \vec{0}_m
    \end{matrix}\right],\\
    B(x(t)) =  \left[\begin{matrix}
    -L_f V(x) - \beta(V(x)) \\
    L_f h(x) + \alpha h(x) \\
    u_{\max}   \\
    u_{\min}
    \end{matrix}\right].
\end{gather*}
Denote the feasible domain of the above CBF-CLF-QP problems as $\Omega_{y}(x(t),\varphi)$ where $\varphi$ represents some related constants and pre-defined parameters. The structured CBF-CLF-QP problems are solved repeatedly at each time step. As a result, the control input $u_{ref}$ is filtered by $u^*$ to ensure safety.

\begin{remark}
    The inefficiency of the CBF-CLF method is revealed in two aspects. On the one hand, the control input, filtered by the sufficiently smooth objective function $f_0$ and continuous time-varying constraints, should be continuous as well. However, CBF-CLF problems \eqref{constraint_ploblem_origin} and \eqref{constraint_ploblem_QP_saturation} are solved as new problems without considering continuity. The lack of utilizing the historical results brings more computational burden for solving independent CBF-QP problems, especially for high dimensional systems with many constraints. On the other hand, the CBF-CLF-QP problems \eqref{constraint_ploblem_QP_saturation} contain myopic objective functions, while the optimization problem, such as tracking a nominal controller, is time-varying, leading to undesirable lag or delay effects. As the nominal dynamics of the systems \eqref{system_dynamic} have been used in CLF and CBF constraints, further exploitation of this information may help reduce or dismiss these bad effects.
\end{remark}

\subsection{Barrier-based Interior Point Method and Prediction}
\begin{assumption}[\bf Convexity and Feasibility]\label{assump_convex_feasible}
For all $ t\ge 0$, the linear constraints $F(y, x(t))$ are convex, and the objective function $f_0$ is uniformly strongly convex such that $\nabla_{yy}f_0 \ge q_c \mathbf{I}_{m_y}$, where $m_y$ is the dimension of $y$. The CBF-CLF-QP problems are always feasible, namely $\Omega_{y}(x(t),\varphi) \notin \emptyset$.
\end{assumption}

\par By introducing the indicator function to punish the infeasible solutions, a constrained problem can be transformed into an unconstrained one with the same optimal solutions \cite{convex_book}. To ensure safety, the barrier method, as a type of interior point method, can be used, which always obtains feasible solutions for optimization. A logarithmic barrier function $\widehat I_{-}(f_i) = - \frac{1}{c} \log(-f_i)$ is introduced to approximate the indicator function $I_{-}(f_i):\Real\to \{0,+\infty\}$, in which $c>0$, $I_{-}(f_i) = 0$ if $f_i < 0$ and $I_{-}(f_i) = +\infty$ otherwise. The logarithmic barrier function is an appropriation of the original constrained problem, as $\lim_{c\to +\infty} \widehat I_{-}(f_i) = I_{-}(f_i)$. The following statements show the suboptimality and quantify the approximation bound of this transformation.

\par Let $F(y, x(t))  = [f_1, \dots, f_p]$ where $p$ is the dimension of $F(y, x(t))$, and $\phi(y,x(t)) = \sum_{i=1}^p -\frac{1}{c}\log(-f_i)$. The constrained problems \eqref{constraint_ploblem_origin} are converted to unconstrained ones
\begin{equation}
    y^* = \min_{y\in \Omega_y(x(t),\varphi)} f_u(y, x, \vartheta) \triangleq  f_0(y, \vartheta) + \phi(y,x). \label{unconstrained_problem}
\end{equation}
\begin{lemma}[Suboptimality \cite{convex_book}]
\label{lemma_suboptimality}
As $c\to \infty$, the barrier-based unconstrained problems \eqref{unconstrained_problem} are the same as the original problems \eqref{constraint_ploblem_origin}. More precisely, problems \eqref{unconstrained_problem} are ${p}/{c}$-suboptimal as
$    \vert f_0(y^*, \vartheta) - f_u(y^*, x, \vartheta)\vert \leq {p}/{c}$.
\end{lemma}
\par Noticing that the optimal solutions are varying according to $x(t)$ and external input $\varphi$ and $\vartheta$, it is rational to include the dynamics of these states into the design of the optimizer. The stationary condition of the optimal solutions hold as $\nabla_y^* f_u(y, x(t), \vartheta) \equiv 0$ whose derivative with respect to $t$ is
\begin{align*}
    \dot \nabla_y^* f_u(y^*, x, \vartheta)
    =&  \nabla_{yy} f_u(y^*, x, \vartheta) \dot y^*\nonumber\\
    & + \nabla_{yx} f_u(y^*, x, \vartheta) \dot x + \nabla_{y\vartheta} f_u(y^*, x, \vartheta) \dot \vartheta.
\end{align*}
The optimal state of $y^*$ and $\dot y^*$ cannot be obtained. Instead, $y$ can be measured with residual error $r \triangleq \nabla_y f_u(y, x, \vartheta)$. As a compromise, the prediction effect is trying to keep the residual error $r$ unchanged according to the evolution of the optimization problem \cite{prediction_correction_unconstrained}. Then, the first-order expansion of the gradient $\nabla_y f_u(y, x, \vartheta)$ is
\begin{align*}
    r =&  \nabla_y f_u(y + \delta y, x + \delta x, \vartheta + \delta \vartheta) 
     \approx  r + \nabla_{yy} f_u(y, x, \vartheta) \delta y\\
     &+ \nabla_{yx}f_u(y, x, \vartheta) \delta x + \nabla_{y\vartheta}f_u(y, x, \vartheta) \delta \vartheta,
\end{align*}
where $\delta a$ stands for the infinitesimal of scalar or vector variables $a$. Subtracting $r$ in both sides, the evolution of $y$ as the optimizer is given by 
\begin{align}
    \dot y =&  - \nabla^{-1}_{yy} f_u(y, x, \vartheta) \Big( \nabla_{yx}f_u(y, x, \vartheta) \dot x \nonumber\\
    &+ \nabla_{y\vartheta}f_u(y, x, \vartheta) \dot \vartheta \Big). \label{predition_interior_point}
\end{align}

\begin{remark}
    The dynamics of $x$ and $\vartheta$ are used for predicting both the evolution of the system state and the control objective. In practice, it is preferred but not necessary to know the gradient of the external input. If the gradient is unknown a priori, the external signal is assumed constant in each step.
\end{remark}

\section{Main Results}
\subsection{Prediction-Correction Safe Controller Design}
In this section, two gradient-based methods are designed to efficiently \textit{correct} the safe controllers, whilst the \textit{prediction} enhancement is considered by combining \eqref{predition_interior_point}. First, the gradient-based descent law with prediction is designed as
\begin{align}
    \dot y = -  & \gamma \nabla_{y} f_u(y, x, \vartheta) - \nabla^{-1}_{yy} f_u(y, x, \vartheta) \Big( \nabla_{yx}f_u(y, x, \vartheta) \dot x  \nonumber\\
    &+ \nabla_{y\vartheta}f_u(y, x, \vartheta) \dot \vartheta \Big), \label{gradient_PClaw_continuous}
\end{align}
where $\gamma > 0$ is the learning rate. The basic idea of this correction is to directly decrease the objective function $f_u$ by its negative gradient, assuring that $f_u$ is non-creasing all the time. Since the gradient-based method will converge to the optimizer for time-invariant problems, it is expected that \eqref{gradient_PClaw_continuous} can find the optimal time-varying solutions by gradient-based correction and Hessian-based prediction. 

\par Notice that the inverse of Hessian $\nabla^{-1}_{yy} f_u(y, x, \vartheta)$ is already computed for prediction, the Newton method can be used for accelerating the convergence process of the approximate solution $y$ to $y^*$ \cite{prediction_correction_interior_point}. The Newton-based prediction-correction dynamics of the approximate optimal solution is given as
\begin{align}
    \dot y = -  & \nabla^{-1}_{yy} f_u(y, x, \vartheta) \Big( \gamma \nabla_{y} f_u(y, x, \vartheta) +  \nabla_{yx}f_u(y, x, \vartheta) \dot x  \nonumber\\
    &+ \nabla_{y\vartheta}f_u(y, x, \vartheta) \dot \vartheta \Big). \label{Newton_PClaw_continuous}
\end{align}
Intuitively, the prediction term keeps the residual error $r$ independent of the evolution of system state $x(t)$ and time-varying immediate objective $f_0$, whilst the gradient method or the Newton method tends to reduce this error $r$. The convergence results of both methods are presented as follows, where all proofs of the results are given in Appendix.

\begin{theorem}
    \label{theorem_convergance_gradient}
    Suppose that Assumption \ref{assump_convex_feasible} holds. The prediction-correction solutions in \eqref{gradient_PClaw_continuous} and \eqref{Newton_PClaw_continuous} will converge to the time-varying optimal solutions $y^*$ of unconstrained problems \eqref{unconstrained_problem} as
    \begin{align}
    \norm{y - y^*}   \leq \frac{\sigma(t_0)}{q_c} e^{-\hat \gamma(t - t_0)},
\end{align}
where $\sigma(t_0) = \norm{\nabla_y f_u (y(t_0), x(t_0), \vartheta(x(t_0),t_0))}$ and $\hat \gamma = q_c \gamma$ for gradient law \eqref{gradient_PClaw_continuous} or $\hat \gamma =  \gamma$ for Newton law \eqref{Newton_PClaw_continuous}.
\end{theorem}

\par It is shown in Theorem \ref{theorem_convergance_gradient} that both gradient-based and Newton-based laws can track the optimal solution exponentially, however at different rates. Note that Newton's law is not necessarily faster than the gradient law in a continuous setting for the prediction-correction method.

\begin{remark}[Choosing Barrier Parameter]
    In the interior point method, the optimal solutions of constrained problems \eqref{constraint_ploblem_origin} are reached by increasingly tuning the barrier parameter $c$ in $\phi(y, x)$ \cite{convex_book}. This operation is explicitly contained in the online time-varying optimization methods in \cite{prediction_correction_interior_point} where $c$ is exponentially increasing such that the evolution of $y$ will converge to the optimal solution $y^*$ of \eqref{constraint_ploblem_origin} exponentially fast. In practice, $c$ should be designed carefully to enhance the robustness \cite{j2006numerical_convex}. The tuning of the barrier parameter $c$ is implicitly contained in $\nabla_{y \vartheta} \dot \vartheta$ in \eqref{gradient_PClaw_continuous} and \eqref{Newton_PClaw_continuous} as a kind of external input in case of complex notations.
\end{remark}

\subsection{Safety and Robustness Analysis}
While the dynamic controller approximates the time-varying optimal solutions of problems \eqref{unconstrained_problem} at an exponential rate, it is not sufficient to guarantee the safety of the systems yet. An illustrative experiment on how an exponential tracking controller may break the safety constraints can be found in \cite{model_free_cbf}. Besides, the ubiquitous disturbances and modeling errors require the algorithms to be robust. In what follows, the safety and robustness of the two prediction-correction methods are studied. For the sake of simplicity, only CBF-QP is considered, such that $y = u \in \Real^m$. It is noted that the extension from CBF-QP to CBF-CLF-QP is trivial.

\begin{assumption}[Bounded Gradient of CBF \cite{model_free_cbf}]
    \label{assumption_gradient_bound_CBF}
    The gradient of the CBF function $h(x)$ defined in Definition \ref{definition_CBF} is finite such that $\forall x \in \mathcal{C}$, $\nabla_x h(x) \leq b_h$ where $b_h > 0$.
\end{assumption}
\begin{theorem}
    \label{theorem_safety}        
    Let Assumptions \ref{assump_convex_feasible}, \ref{assumption_gradient_bound_CBF} hold and $c\to \infty$, which means the optimal solutions of \eqref{unconstrained_problem} are the same as the optimal solutions of \eqref{constraint_ploblem_origin}. Denote these optimal solutions as $u^*$ uniformly. Then, both the gradient-based and the Newton-based methods ensure the safety of the systems provided that $ \hat \gamma > \alpha$, $\alpha^\prime \triangleq q_c \left(  \hat\gamma - \alpha\right) / b_h$ and  $(x(t_0), y(t_0)) \in \mathcal{C}_N$, where $\mathcal{C}_N \triangleq \left\{ \left(x, y\right) | h_N(x, y) \ge 0 \right\}$ and
    \begin{equation}
        h_{N}(x, y) = -\norm{\nabla_y f_u (y, x, \vartheta)} + \alpha^\prime h(x). \label{threshold_CBF_safety}
    \end{equation}
\end{theorem}

\par From the results in Theorem \ref{theorem_convergance_gradient}, safety can be ensured when the convergence rate of the approximate solution $u$ to the optimal $u^*$ is higher than the pre-defined allowable rate of change in CBF constraints. This is consistent with the conditions of a safe tracking controller in \cite{model_free_cbf}. The following statement presents the safety criteria for bounded $c$ and the biased approximate solutions.

\begin{corollary}
\label{colollary_safe_constant_c}
Let Assumption \ref{assump_convex_feasible}, \ref{assumption_gradient_bound_CBF} hold and $c$ be constant or time-varying but bounded. Define the approximate optimal solutions of problem \eqref{unconstrained_problem} as $\widehat u^*$. Both the gradient-based and the Newton-based methods ensure the safety of the systems provided that $\hat \gamma > \alpha$, $\alpha^\prime \triangleq q_c \left( \hat\gamma - \alpha\right) / b_h$ and  $(x(t_0), y(t_0)) \in {\mathcal{C}}_N$ with ${\mathcal{C}}_N$ defined in Theorem \ref{theorem_safety}.
\end{corollary}

\par The proof of Corollary \ref{colollary_safe_constant_c} is the same as that of Theorem \ref{theorem_safety} with the fact that the estimated solutions are always in the interior of the feasible set thanks to the idea of interior point optimization. Therefore, the CBF constraints \eqref{constraint_ploblem_origin_cbf} can always be met by the approximate optimal solutions $\widehat u^*$. The proof is omitted for brevity.

\par So far, gradient-based and Newton-based dynamics can both perform well in an ideal environment where no disturbance or inaccuracy exists in the learning progress, such that the models are perfectly known, and all computations are accurate. To analyze the robustness of the methods, the disturbances in the derivatives of prediction parts including the model uncertainty and random noises are considered. The term of prediction in both \eqref{gradient_PClaw_continuous} and \eqref{Newton_PClaw_continuous} needs the information of dynamics as $\dot x$ and $\dot \vartheta$. It is assumed that disturbances exist in obtaining this information, as described below.
\begin{equation}
    \norm{\nabla_{yt}f_u(y, x, \vartheta) - \widehat{\nabla_{yt}f_u}(y, x, \vartheta)} \leq \sup_{t > t_0} \norm{\varrho(t)}, \label{disturbances_of_gradient_prediction}
\end{equation}
where $\varrho(t) \in \Real^{m_y}$ represents the estimation error, and $\widehat{\nabla_{yt}f_u}(y, x, \vartheta)$ denotes the utilized estimation of the information of origin dynamics $\nabla_{yt}f_u(y, x, \vartheta) = \nabla_{yx}f_u(y, x, \vartheta) \dot x + \nabla_{y\vartheta}f_u(y, x, \vartheta) \dot \vartheta$. The bound \eqref{disturbances_of_gradient_prediction} means that the estimation is in an admissible compact set of accurate dynamics. 
\begin{theorem}
    \label{theorem_ISS_with_disturbances}
    Suppose that Assumption \ref{assump_convex_feasible} holds. The prediction-correction systems \eqref{gradient_PClaw_continuous} and \eqref{Newton_PClaw_continuous} are both input-to-state stable (ISS) and the approximate error is ultimately bounded as
    \begin{equation}
        \lim_{t\to \infty}\norm{y - y^*} \leq \frac{\sup_{t>t_0}\norm{\varrho(t)}}{\xi q_c}
    \end{equation}
    for $\xi \in (0,1)$, and the convergence rate reduces to $\hat \gamma(1 - \xi)$.
\end{theorem}

Theorem \ref{theorem_ISS_with_disturbances} has demonstrated the robustness of the prediction-correction systems. In CBF-based safety-critical control, the evolution of the approximate errors should be also important to ensure robust safety, as $\frac{d}{dt}\norm{\nabla_y f_u}$ in Theorem \ref{theorem_safety} and Corollary \ref{colollary_safe_constant_c}. The evolution can be quantitatively described in virtue of the modified Lyapunov function $V_M(\sigma(t), t) = \sqrt{2 V(\sigma(t), t)}$, whose derivative is ${\dot V(\sigma(t), t)}/{\sqrt{2 V(\sigma(t), t)}}$. Following \eqref{lyapunov_derivative_temp}, one has
\begin{equation}
        \dot V_M(\sigma(t), t) \leq - \hat \gamma V_M(\sigma(t), t) + \sup \norm{\varrho(t)}. \label{quantitative_describe_evolution}
\end{equation}
Notice that $V_M(\sigma(t), t) = \norm{\nabla_y f_u (y, x, \vartheta)}$, the evolution $\frac{d}{dt}\norm{\nabla_y f_u (y, x, \vartheta)}$ can be described by \eqref{quantitative_describe_evolution}. To ensure safety, a robust CBF condition is addressed to include the worst-case disturbances $\sup\norm{L_d h(x)}$ as
\begin{align}
     L_f \widetilde h(x) + L_g \widetilde h(x) u &+ \alpha \widetilde h(x) + \sup\vert L_d \widetilde h(x)\vert \ge 0, \label{disturbances_cbf}
\end{align}
replacing \eqref{constraint_ploblem_origin_cbf} in the original problems \eqref{constraint_ploblem_origin}. An enlarged invariant set is defined by $\widetilde{\mathcal{C}} = \left\{ x\vert \widetilde h(x)  \ge 0 \right\}$. The following statements give the safety criteria of the synthetic controllers.
\begin{corollary}
    \label{colollary_safe_with_disturbances}
Let Assumption \ref{assump_convex_feasible}, \ref{assumption_gradient_bound_CBF} hold. Define the approximate optimal solutions of problems \eqref{unconstrained_problem} as $\widehat u^*$. The gradient- and Newton-based methods ensure the safety of the systems under disturbances in \eqref{disturbances_of_gradient_prediction} provided that $(x(t_0), y(t_0)) \in \widetilde{\mathcal{C}}_N$, $\widetilde{\mathcal{C}}_N \triangleq \left\{ \left(x, y\right) | \widetilde h_N(x, y) \ge 0 \right\}$ where
    \begin{equation*}
       \widetilde h_{N}(x, y) = -\norm{\nabla_y f_u (y, x, \vartheta)} + \alpha^\prime  \widetilde h(x) + \frac{1}{\alpha} \sup\norm{\varrho(t)} .
    \end{equation*}
\end{corollary}

\begin{remark}[Design of CBFs]
    The conditions for Lipschitz continuity of the optimal controller $u^*$ are analyzed in \cite{cbfdefine} which is a critical assumption to guarantee the feasibility of the proposed methods. When the relative degrees of barrier function $h(x)$ is higher than $1$, the exponential CBFs are effective to ensure safety as described in \cite{exponential_CBF}, which will be utilized in the simulations part of this paper. It is always hard to determine the appropriate CBFs for the control invariant set, which can significantly affect the performance of the CBF methods as well as the prediction-correction methods as shown in \cite{backup_cbf}, where a backup policy is introduced to tackle this issue. Besides, the Hamilton-based computation for reachability may help to improve the CBF methods \cite{HJ_reachability,cbf_reachibility}.
\end{remark}
\begin{remark}[Comparison of Two Methods]
    Although it is shown in Theorem \ref{theorem_convergance_gradient} that the minimum convergence rates of both dynamics are almost the same, the convergence rate of the gradient-based method in \eqref{gradient_convergence_rate} can be much larger than that of Newton-based method in \eqref{newton_convergence_rate} when $c$ is initialized small or the solutions approach to the constraint boundaries, namely when $\sum_{i = 1}^p A_i(x)/ c$ is large. It can also increase the robustness as seen in \eqref{lyapunov_derivative_temp} when considering the term $\nabla_{yy} \phi(x, y)$. Note that Newton-based descent law is appropriate for the originally unconstrained problems, while the barrier-based unconstrained problem \eqref{unconstrained_problem} with ill-conditioned Hessian of $f_u(y,x,\vartheta)$, which is needed in the prediction terms, would not benefit the control process. The exploration of the Hessian of $f_0(y, \varepsilon)$ to accelerate the control process can be a promising task for efficient constrained optimization \cite{j2006numerical_convex}. In addition, the gradient-based law can work well without prediction term if the evolution of system dynamics and external input is slow, then $\nabla_{x}\sigma(t)\dot x + \nabla_{\vartheta}\sigma(t) \dot \vartheta$ can be viewed as disturbances.
\end{remark}

\subsection{Online Implementation}
The continuous prediction-correction dynamics \eqref{gradient_PClaw_continuous} and \eqref{Newton_PClaw_continuous} work perfectly under high and fast computing power, which can hardly achieve in practice. In addition to the continuity, the inverse calculation takes time to ensure accuracy, and the predictive evolution needs more prior information about the systems as well as the external input. Therefore, the discrete versions with less prior information are preferred, given as:
\begin{align}
    \text{Gradient: }~ y_{k+1} = ~ & y_{k} - h \gamma G_k - h H_k^{-1} \widehat{\nabla_{yt}f_k} ,\label{discrete_dynamics_gradient}\\
    \text{Newton: }~ y_{k+1} = ~ & y_{k}  - h H_k^{-1} \left( \gamma G_k +  \widehat{\nabla_{yt}f_k}\right) \label{discrete_dynamics_newton},
\end{align}
where $h$ is the sampling step, $k$ represents the sampling series, $G_k \triangleq  \nabla_{y}f_u(y_k, x_k, \vartheta_k) $, $H_k \triangleq \nabla^{-1}_{yy} f_u(y_k, x_k, \vartheta_k)$ and $\widehat{\nabla_{yt}f_k}$ is the estimation of $\nabla_{yt} f_u(y_k, x_k,\vartheta_k)$ in the prediction terms satisfying \eqref{disturbances_of_gradient_prediction}. The controller is implemented by zero-order hold (ZOH). A promising structure of $\widehat{\nabla_{yt}f_k}$ can be designed by $ (G_k - G_{k - 1}) / h$. Let $a_i(x) \in \Real^{m_y}$ be the $i$th column of $A(x)$ and $b_i(x)$ be the $i$th element in $B(x)$ in \eqref{constraint_ploblem_QP_constraints}, $i = 1,\dots,p$. The compact forms of the derivative and Hessian of $f_u(y, x, \vartheta)$ are given as
\begin{align*}
        \nabla_{y} f_u(y, x, c) = & \nabla_{y} f_0(y,\vartheta)  + \sum_{i=1}^p \frac{1}{c f_i} a_i(x)\\
        = & 2Q y - 2 H(x)  +  c^{-1}A(x) d(y, x),\\
        \nabla_{yy} f_u(y, x, c) = & \nabla_{yy} f_0(y,\vartheta) +  \sum_{i=1}^p \frac{1}{c f_i^2} a_i(x) a_i(x)^\Tran \\
        =&  2Q  + c^{-1}A(x) D(y, x) A^\Tran(x),
\end{align*}
where $ d(y,x) = \left[d_1(y,x), \dots, d_p(y,x) \right]^\Tran $ and $D(y,x) = diag\{ d^2_1(y,x), \dots , d^2_p(y,x)\}$ with $d_i(y,x) = f^{-1}_i(y,x)$, $i = 1,\dots,p$. Therefore, the explicit expressions of $G_k$ and $H_k$ in \eqref{discrete_dynamics_gradient} and \eqref{discrete_dynamics_newton} are
\begin{align}
        G_k = 2Q y_k - 2 H(x_k)  +  c^{-1}A(x_k) d(y_k, x_k), \label{calculation_Gk}\\
        H_k = 2Q  + c^{-1}A(x_k) D(y_k, x_k) A^\Tran(x_k).\label{calculation_Hk}
\end{align}

\par The stability of the gradient-free discrete dynamics for unconstrained time-varying convex optimization has been studied in \cite{prediction_correction_unconstrained}, in which the principles of choosing the learning rate $\gamma$ to adapt the sampling period $h$ were designed for bounded approximate errors. However, for the barrier-function based unconstrained problem \eqref{unconstrained_problem}, the assumption on the boundedness of the derivatives of sufficiently smooth objective functions $f_u$ in \cite{prediction_correction_unconstrained} is not satisfied because the derivative of the logarithmic barrier function can tend to infinite as $y$ approach to the surface of $\omega_y (x(t), \varphi)$, such that $f_i \to 0$. It is even harder to set $c \to \infty$ for pursuing the optimal solutions as designed in \cite{prediction_correction_interior_point}. More important thing is, the discrete version can lead to the break of safety restrictions easier when $c$ gets larger.

\par It becomes tractable when the expectation of the optimal solution is reduced, such that $c$ is bounded rather than tending to infinity. In this situation, the approximate errors are given in Lemma \ref{lemma_suboptimality}. A good property of this suboptimality is that the global principles of the boundedness can be reduced to local ones, which is quite easy to achieve since $f_i = 0$ will not occur through the interior point optimization. More specifically, the optimal solutions of the unconstrained problem \eqref{unconstrained_problem} with bounded $c$ satisfying $ \nabla_y \phi(y^*)= -\nabla_y f_0(y^*)$ while $ \nabla_y \phi(y^*) \to -\infty$ if $f_i \to 0$. Therefore, the optimal solutions lie in a compact subset of $\mathcal{C}$ and the derivatives of $f_u(y, x, \vartheta)$ satisfy the bounded assumptions in \cite{prediction_correction_unconstrained}, which legalizes the conditioned discretization. 

\begin{remark}[Relaxation on Inverse of Hessian]
    The calculation of the inverse of Hessian brings undesirable computational costs when using the presented methods. This can be avoided when a computation affordable projection operation is admissible. The sampled-data-based first-order prediction method relying on only the Hessian of the original objective function has been proposed in \cite{prediction_correction_constrained_projection} with a projection operator to ensure feasibility. More effort can be made to exploit the structure of CBF-CLF-QP, where a fast computation law of the inverse Hessian matrix can be created. For instance, one can rewrite \eqref{calculation_Hk} as
    \begin{equation}
        H_k = 2Q  + c^{-1}A(x_k) D^{\frac{1}{2}} D^{\frac{1}{2}} A^\Tran(x_k) = K_Q + L_c^T L_c,\label{calculation_Hk_decomposed}
    \end{equation}
    where $K_Q = 2 Q$ and $L_c = A(x_k) D^{\frac{1}{2}} / \sqrt{c}$. When $L_c$ is sparse, which is common when a system is underactuated or of higher DOF, the calculation of the inverse matrix can be greatly saved as studied in \cite{0021B10}.
\end{remark}
\begin{remark}[Certificate the Feasibility]
    In the discrete version of prediction-correction methods, the control input or approximate solutions may leave the invariant set for big step size or large barrier parameter $c$. An economical way to avoid it is to check the feasibility of the solutions before each update. If there is any broken constraint, the step is rejected and a new, shorter trial step is used for another iteration, such as line search and trust-region method \cite{j2006numerical_convex,convex_book}.
\end{remark}


\section{Simulations}

\par \emph{Example 1.} Consider a two-dimensional integrator system as $\dot x = v$ with $x = [x_1, x_2]^\Tran$ and $v = [v_1, v_2]^\Tran$ the position and velocity, respectively. The control goal is to move the position $x$ to the desired position $x_d$ by designing the velocity input, whilst the position should stay away from some unsafe areas. In this example, the unsafe region is described by circles with uniform radius $r$ and different center points $x_c$. As discussed in \cite{model_free_cbf}, the CBF function can be defined as
\begin{equation}
    h(x) = \norm{x - x_c} - r. \label{example1_CBF}
\end{equation}
Note that $h(x)>0$ implies that the position $x$ is not in the unsafe region. To further simplify the problem, the nominal velocity steering system state to $x_c$ is designed by a feedback law $v_d = -k_d(x - x_d)$ with $k_d>0$. Then, a CBF-QP can be constructed to filter the nominal velocity input to a safe one as follows:
\begin{gather}
    v^* = \min_{v \in \Real^{2}} v^\Tran v - 2 v_d^\Tran v, \label{example_avoid_object}\\
    s.t. ~ -\nabla_{x} h(x) v - \alpha (h(x)) \leq 0, \nonumber
\end{gather}
where $v^*$ is the filtered safe input. The gradient of $h(x)$ can be calculated as ${(x - x_c)}/{\norm{x - x_c}}\triangleq n_c$. Based on the Karush-Kuhn-Tucker (KKT) conditions, the explicit optimal safe solution of \eqref{example_avoid_object} takes the following form:
\begin{equation}
    v = v_d + \max \left\{ -n_c^\Tran v_d - \alpha(h(x)), 0 \right\} n_c. \label{example1_optimal_solution}
\end{equation}

\par The initial system position is $[0.0, 0.5]^\Tran$, and the target position is $[2.5, 3.0]^\Tran$. To move the position approaching the target, the feedback coefficient is set as $k_d = 1.1$. The single unsafe region is determined by $x_c = [1.0, 1.0]^\Tran$ and $r = 0.8$. Notice that the feedback input will force the system state into the unsafe region. To avoid this, the velocity input is filtered by \eqref{example1_optimal_solution} where $\alpha = 4.0$. Alternatively, the gradient- or Newton-based dynamical laws \eqref{gradient_PClaw_continuous} and \eqref{Newton_PClaw_continuous} can be utilized to approximate the optimal safe input policy. The parameters in the approximation laws are set as $\gamma = 15.5$, $c = 1.1 e^{0.9t}$, discrete step $dt = 0.001s$. The barrier parameter is set monotonically increasing to reduce the error bound as explained in the interior point method. It is noted that the bound of increasing rate is related to the step-size $dt$. The control input between each sampling time is kept by ZOH and all the gradients and Hessian matrices are known, which are given in Appendix \eqref{appendix_example1}.

\par
\begin{figure}[tb]
    \centering
	\subfigure[]{
		\centering
        \scalebox{.55}{\includegraphics{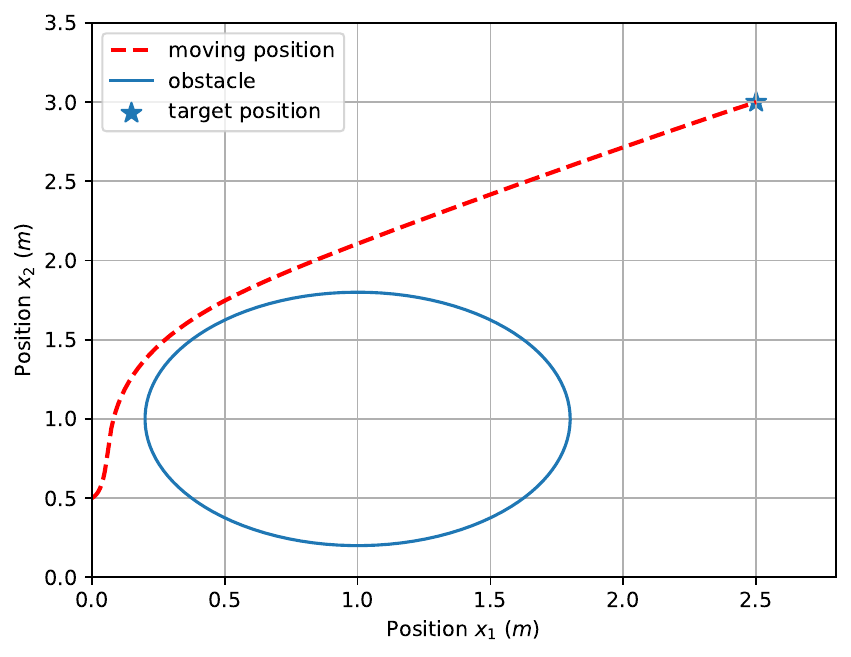}}
		\label{fig_example1_position_gradient}
		}
	\subfigure[]{
		\centering
        \scalebox{.55}{\includegraphics{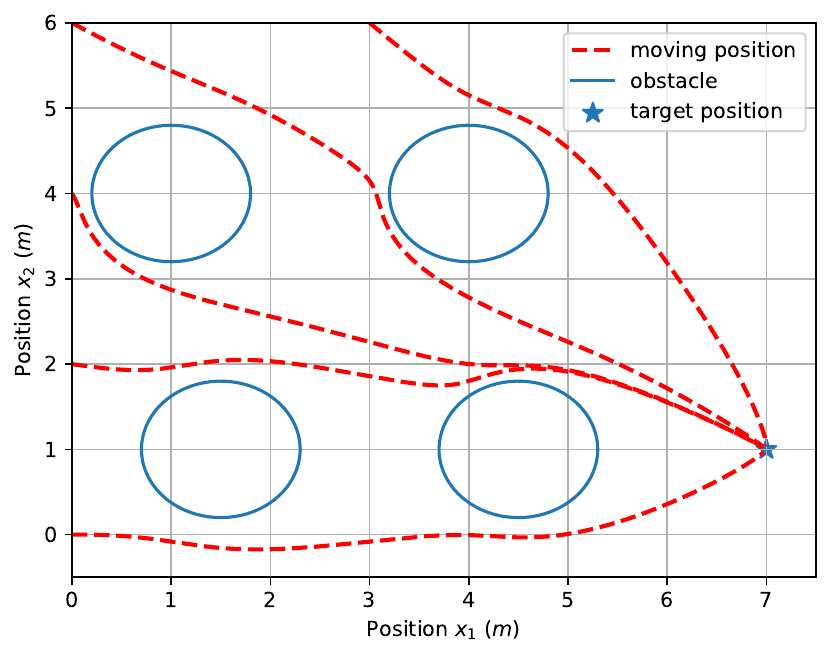}}
		\label{fig_example1_position_newton}
		}
\caption{Position trajectories controlled by \eqref{gradient_PClaw_continuous} and \eqref{Newton_PClaw_continuous} to avoid contact with the unsafe areas. (a) Trajectory under gradient-based input \eqref{gradient_PClaw_continuous} with a single obstacle. (b) Trajectories under Newton-based input \eqref{Newton_PClaw_continuous} from different initial positions with multiple obstacles.}
\label{fig_example1_position}
\end{figure}

\begin{figure}[tb]
    \centering
	\subfigure[]{
		\centering
		\includegraphics[width=4.1cm]{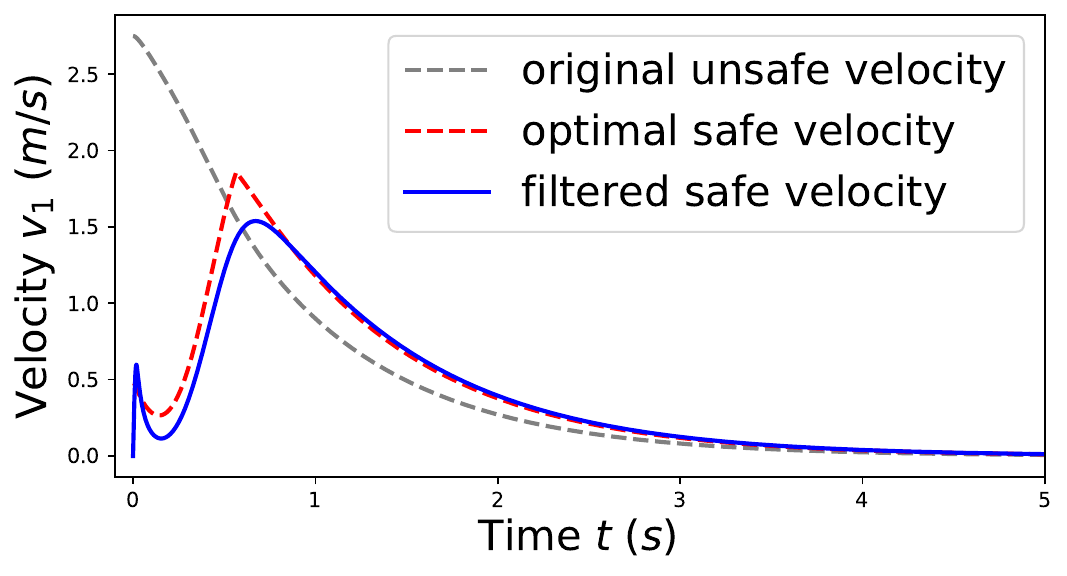}
		\label{fig_example1_single_v1_gradient}
		}
	\subfigure[]{
		\centering
		\includegraphics[width=4.1cm]{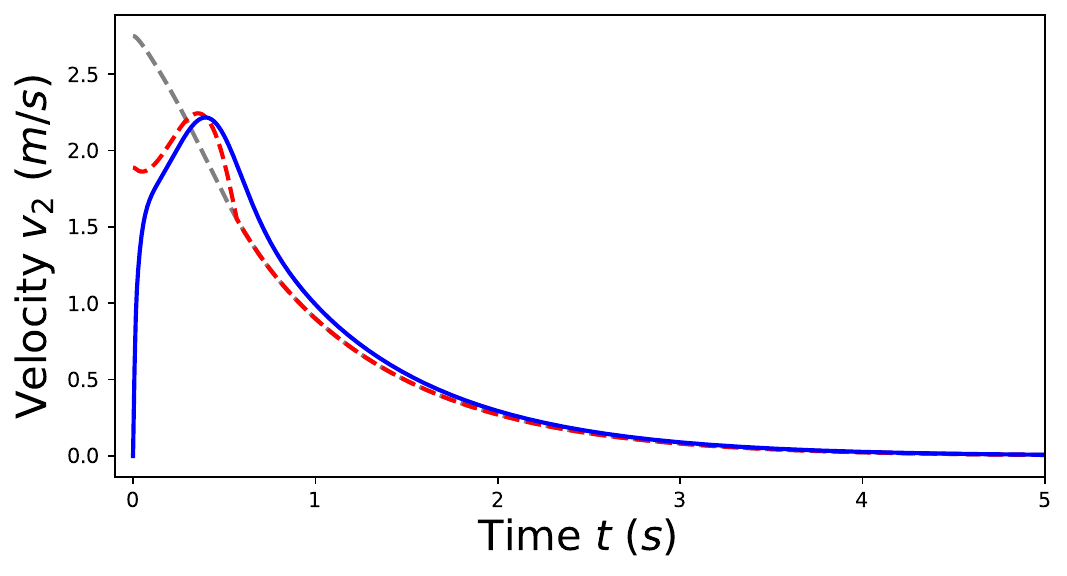}
		\label{fig_example1_single_v2_gradient}
		}
	\subfigure[]{
		\centering
		\includegraphics[width=4.1cm]{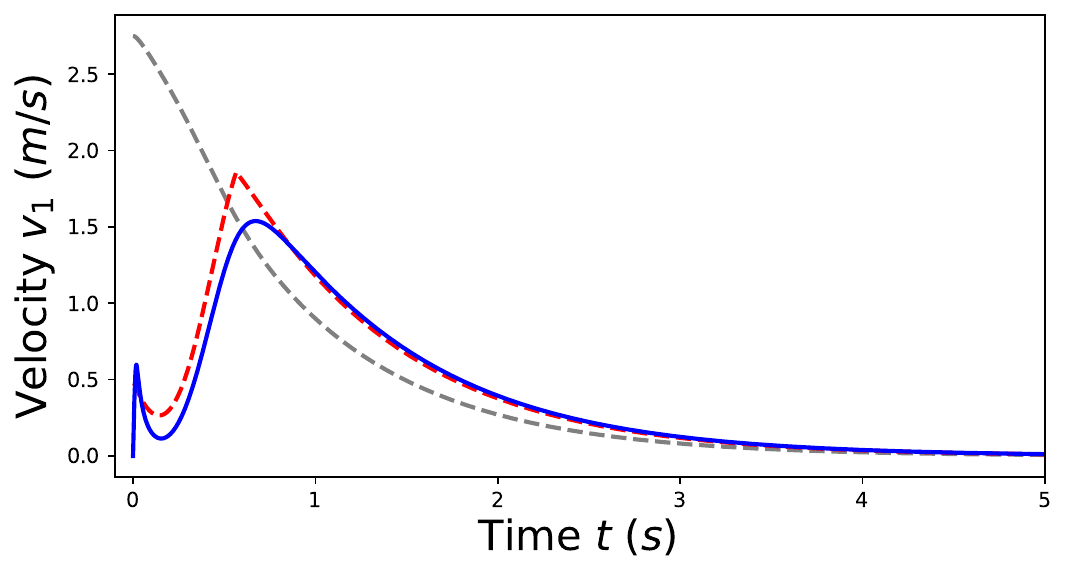}
		\label{fig_example1_single_v1_newton}
		}
	\subfigure[]{
		\centering
		\includegraphics[width=4.1cm]{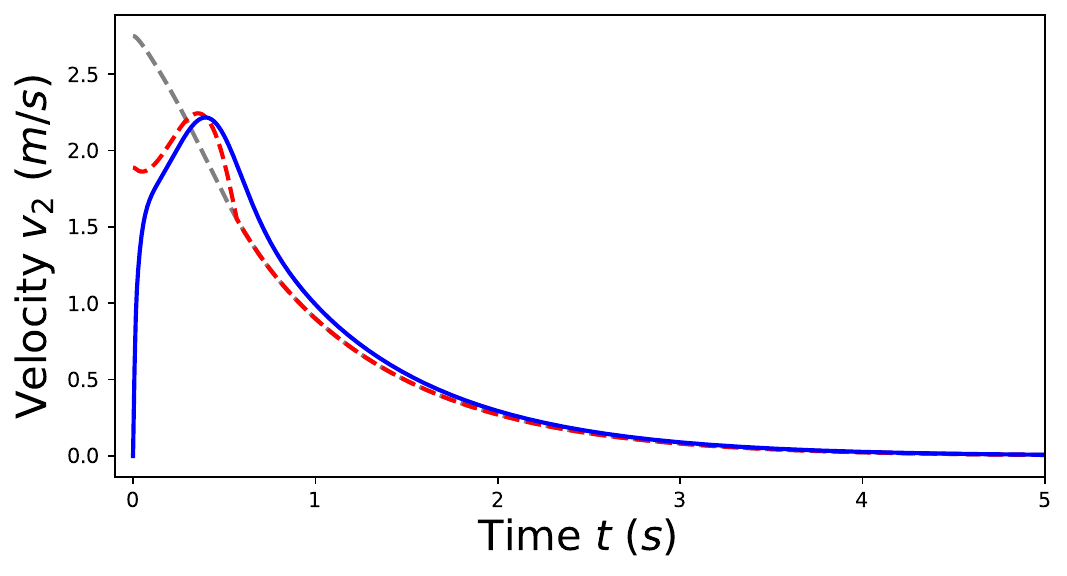}
		\label{fig_example1_single_v2_newton}
		}
\caption{Velocities of feedback law $v_{d}$, optimal CBF-QP based input $v^*$ and approximate optimal safe input $v$ with single obstacle and (a-b) gradient-based law \eqref{gradient_PClaw_continuous}, (c-d) Newton-based law \eqref{Newton_PClaw_continuous}.}
\label{fig_example1_single_gradient}
\end{figure}

 Initially, the velocity is zero and the system state is in the forward invariant set $\{x \vert h(x)>0\}$. The dynamical position generated by the gradient-based method \eqref{gradient_PClaw_continuous} is shown in Fig.~\ref{fig_example1_position}, as which the omitted newton-based result is roughly the same. It is shown that the unsafe region is avoided when moving to the target position. The comparison of the original, optimal, and approximate optimal velocity inputs are shown in Fig.~\ref{fig_example1_single_gradient}, where the optimal input is generated at each step based on the current position and the explicit solution \eqref{example1_optimal_solution}. The results reveal that the descent-based approximate optimal safe control laws can not only track the optimal policy but can also ensure safety during the tracking process. Note that the tracking errors converge to zero because the barrier parameter $c$ is exponentially increasing during the process. It is shown that the gradient-based method has a faster transient response when the tracking error is large, while the Newton-based method performs better when the tracking error is smaller, owing to the discrete sampling and finite barrier parameter $c$. It can be improved when the sampling period is smaller.

\par The safe tracking effectiveness with multiple obstacles is tested in simulations as well. Unlike \cite{model_free_cbf} which only took the nearest obstacle into account, the CBF constraints of different obstacles are all considered in the descent-based methods. The result using the Newton-based method \eqref{Newton_PClaw_continuous} is given in Fig.~\ref{fig_example1_position_newton}, which is roughly the same as the gradient-based method as well. The parameters are set as $k_d = 0.2$, $c = 0.9 e^{0.2t}$, discrete step $dt = 0.01s$. Other parameters remain the same as those with a single obstacle. Five different initial positions and four obstacles are considered, where the initial points are $[0.0, 2 k]^\Tran$ and $[3.0, 6.0]^\Tran$, $k = 0,1,2,3$, and the center points of obstacles are $[1 + m, 4]^\Tran$, and $[1.5 + m, 1]^\Tran$, $m= 0, 3$. It is shown that the tracking control is successfully filtered to force the system to stay in the safe region and ultimately reach the control target. 
Furthermore, the average computing time of the two methods is $0.25ms$ for single obstacles and $0.92ms$ for multiple obstacles, implying that the calculation burden is low for obtaining the approximate optimal solutions.




\begin{figure}[tb]
\centerline{
        \scalebox{.6}{\includegraphics{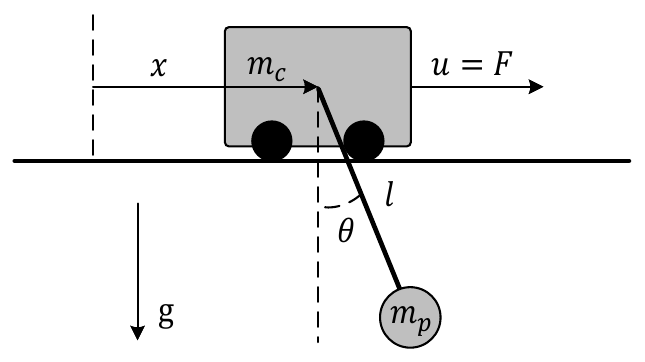}}
    }
\caption{The structure of a 2-D cart-pole system.}
\label{fig_example2_cartpole}
\end{figure}

\begin{figure*}[tb]
    \centering
    \scalebox{.55}{\includegraphics{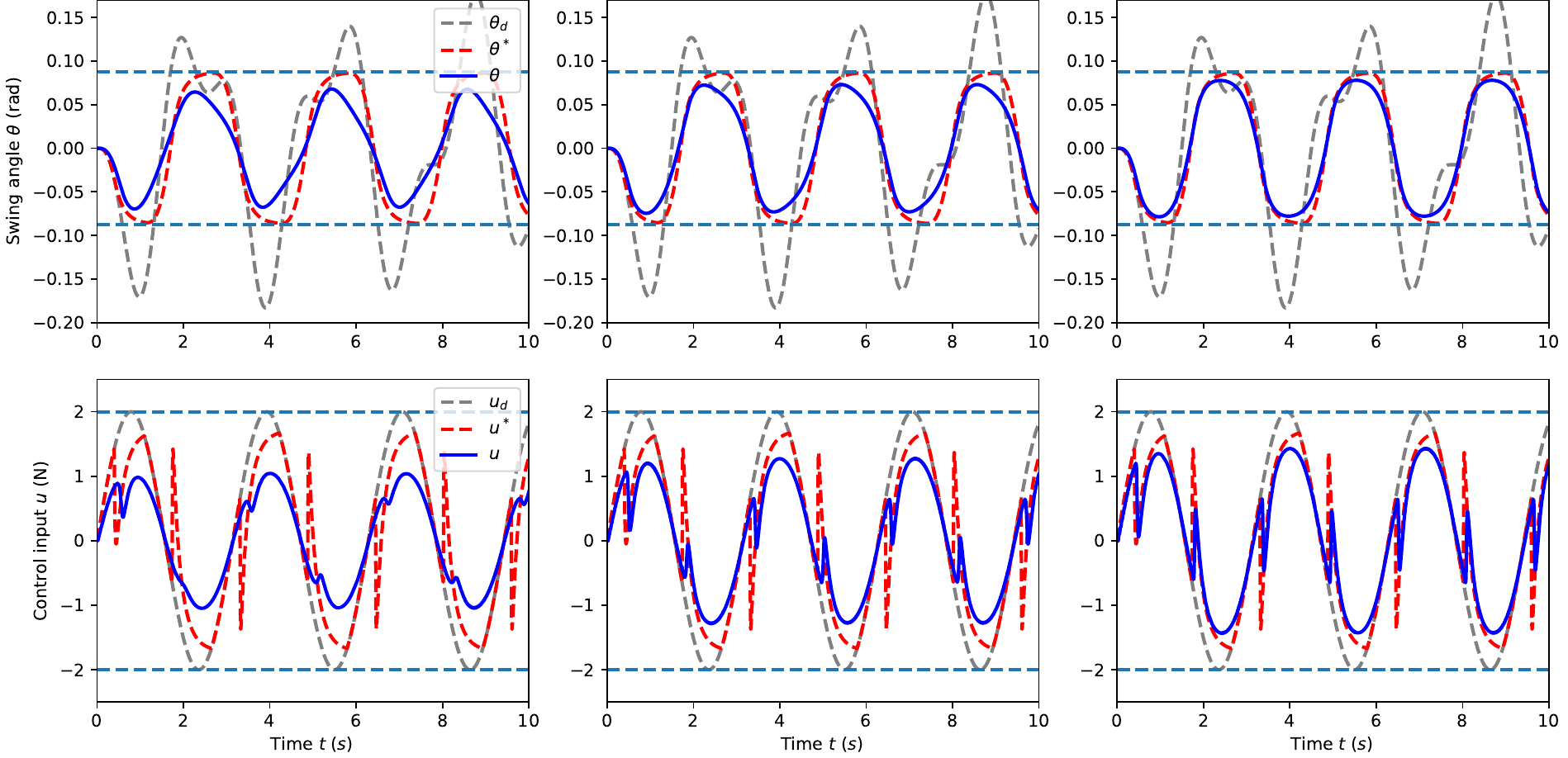}}
    \caption{Trajectories of the swing angle and control input of different settings. The barrier parameter $c$ is different in three columns, from (Left) $0.5$, (Middle) $1.0$, to (Right) $2.0$. In the first row, the gray dotted, red dotted and blue lines represent the angles controlled by original input $u_d$, optimal safe input $u^*$, and gradient-based safe input $u$. The horizontal dotted lines represent the maximum and minimum admissible swing angles. In the second row, the gray dotted, red dotted, and blue lines are generated by the original given input $u_d$, the safe input $u^*$ filtered by CBF-QP and the safe input $u$ filtered by gradient-based algorithm \eqref{gradient_PClaw_continuous}. The horizontal dotted lines represent the maximum and minimum admissible control inputs.
    }
    \label{fig:antiswing_total}
\end{figure*}



\par \emph{Example 2.} 
In this example, an anti-swing control task for a cart-pole system is simulated to show the effectiveness from a more practical perspective. As shown in Fig.~\ref{fig_example2_cartpole}, the dynamics of the cart-pole system is given as
\begin{align}
    \dot q = \left[\begin{matrix}
    \dot x \\
    \dot v   \\
    \dot \theta\\
    \dot \omega
    \end{matrix}\right] =  \left[\begin{matrix}
    v\\
    f_{v}(\theta, \omega) \\
    \omega\\
    f_{\omega}(\theta, \omega)
    \end{matrix}\right] + \left[\begin{matrix}
    0 \\
    g_v(\theta, \omega)   \\
    0\\
    g_\omega(\theta, \omega)
    \end{matrix}\right] u, \label{cart_pole_system_dynamics}
\end{align}
where $m_c$, $m_p$ are the mass of the cart and the mass of the pole respectively, $x$ is the horizontal position of the cart, $u$ is the external force input, $\theta$ and $l$ are the angle and length of the pendulum (assume the mass of the pendulum is small enough), $v$ and $\omega$ are the horizontal and angular velocities, $f_v$, $f_\omega$, $g_v$ and $g_\omega$ are given in Appendix \ref{appendix_example2}. The control task is trying to follow the given input signal $u_d$ regulating this system as much as possible without sharply swinging, in other words, keeping the angle $\theta$ in the admissible range. Let the admissible range be $[-r,r]$, $r > 0$. Then, the candidate CBF can be chosen as $h = r^2 - \theta^2$, whose derivative is $ \dot h = -2 \theta \omega$.

Notice that the cart-pole system is the one with more than one DOF. It will show that the presented methods apply to these systems as well. In the spirit of exponential CBF in \cite{exponential_CBF}, define the following function as an alternative CBF: $h_e = \dot h + \mu h \ge 0$, $\mu > 0$, whose derivative is
\begin{align*}
    \dot h_e = & \ddot h + \mu \dot h \\
    = & -2\theta \left(f_\omega(\theta, \omega) + g_\omega(\theta, \omega) u \right) - 2 \omega^2 - 2\mu \theta\omega\\
    \ge&  -\alpha \left( -2\theta\omega + \mu (r^2 - \theta^2) \right),
\end{align*}
being the linear constraint of $u$. Furthermore, the input saturation is also considered by $u_{\min}$ and $u_{\max}$. In all, the standard form of CBF and saturation constraints are $A^\Tran(q) u - B(q) \leq 0$, where $A(q)$ and $B(q)$ are given in Appendix \ref{appendix_example2}.

\par The system parameters are set as $m_c = 1kg$, $m_p = 1kg$, $l = 2.0m$ and gravitational acceleration $g = 9.8 m / s^2$. The parameters of CBF-based QP are set to be $\alpha = 7.5$, $\mu = 7.5$, $r = 5^\circ$, $u_{\min} = -3.0 N$ and $u_{\max} = 3.0 N$. The discrete version of the prediction-correction gradient-based law \eqref{discrete_dynamics_gradient} is used to efficiently obtain the approximate optimal safe solution, where the parameters are designed as $\gamma = 20$ and sampling step $dt = 0.001s$. Note that the sampling step is based on the maximum average computation time for solving CBF-QP. As for the barrier parameter, three different values, $c \in \left\{0.5, 1.0, 2.0\right\}$ are tested to illustrate the suboptimality in Lemma \ref{lemma_suboptimality}. Assume that the time derivative ${\nabla_{yt}f_k}$ is not known and is replaced by $ \left(G_k - G_{k - 1}\right) / dt$ in \eqref{calculation_Gk}, which can be viewed as causing an unknown but bounded disturbance. The boundedness is checked at each sampling step by $\norm{G_k - G_{k-1}} \leq D_d$ and $G_d = 0.01$ is the pre-determined threshold. If the threshold is exceeded, a smaller step can be used to correct it. For simplicity, in this example, the prediction term is abandoned and only the gradient term is used to track the time-varying suboptimal solutions. It is tested that this algorithm has desirable numerical stability.

\par Initially, system state $x$, $\theta$, $v$, and $\omega$ are all set zero. The given input signal is manually designed as $u_d(t) = 4\sin{t}\cos{t}$ which does not break the saturation constraints. However, forced by $u_d$, the swing angle denoted by $\theta_d$ may become large and exceed the admissible range, shown as the gray dotted line in the first row of Fig.~\ref{fig:antiswing_total}. Therefore, the safe filter is necessary for the anti-swing control task. The optimal safe solutions $u^*$ in the sense of the CBF method can be obtained at each step by repeatedly calling the QP solver. Such solutions replace $u_d$ as the input to the cart-pole system \eqref{cart_pole_system_dynamics}. The control result is shown as the red dotted line in Fig.~\ref{fig:antiswing_total}, which implies that the swing angle $\theta^*$ is always in the safe range and will contact the safe boundary periodically. Then, the suboptimal solutions are obtained by the discrete gradient-based prediction-correction method, where the result is shown as the blue line in Fig.~\ref{fig:antiswing_total}. Similarly, the swing angle is in the safe range, but without touching the boundaries. This is because the finite barrier function $c$ makes the solutions suboptimal. Since the stability and safety of the suboptimal methods have been analyzed, one advantage of the presented methods is the robustness compared to exact CBF-QP methods in the presence of uncertainty.

\par The filter-based control input is given in the second row of Fig.~\ref{fig:antiswing_total}. It is seen that all three control inputs satisfy the saturation constraints. The optimal safe controller filtered by the CBF-QP method has some impulsive discontinuous points, requiring the actuator to be sensitive enough to track the safe input signals. However, since the boundary is touched under the CBF-QP method, it is likely to break the safety constraints when the actuator cannot switch its output. When considering the suboptimal solutions, a more smooth trajectory is obtained as the blue line in Fig.~\ref{fig:antiswing_total}. It can perform better when the actuator must be tuned smoothly.

\begin{table}[tb]
    \centering
    \caption{Numerical results of experiments on anti-swing control.}
    \begin{tabular}{|c|c|c|}
     \hline
     Method & Performance & Average Computation Time\\
     \hline
     QP (CVXOPT \cite{CVXPY_code}) & $3.872$  & $1.380 (\pm 0.416) \mathrm{ms}$   \\
     \hline
     {PCL \eqref{discrete_dynamics_gradient}, $c=0.5$} & $6.138$  & $0.485 (\pm 0.029) \mathrm{ms}$ \\
     \hline
     {PCL \eqref{discrete_dynamics_gradient}, $c=1.0$} & $5.171$     & $0.564 (\pm 0.153) \mathrm{ms}$ \\
     \hline
     {PCL \eqref{discrete_dynamics_gradient}, $c=2.0$} & $4.526$     & $0.496 (\pm 0.073)    \mathrm{ms}$ \\
     \hline
     {PCL \eqref{discrete_dynamics_newton}, $c=0.5$} & 6.$216$     & $0.440 (\pm 0.034)    \mathrm{ms}$ \\
     \hline
     {PCL \eqref{discrete_dynamics_newton}, $c=1.0$} & 5.$319$     & $0.397 (\pm 0.022)    \mathrm{ms}$ \\
     \hline
     {PCL \eqref{discrete_dynamics_newton}, $c=2.0$} & 4.$765$     & $0.396 (\pm 0.017)    \mathrm{ms} $ \\
     \hline
    \end{tabular}
    \label{tab:computation_time}
\end{table}


In terms of efficiency, the numerical results of the average computation time of each method with different $c$ are shown in Table \ref{tab:computation_time}, where $5$ repeated experiments are conducted with performance, mean values, and standard deviations of the average computation time at each discrete step. The performance computes the sum of the myopic objective values in all steps. As the barrier parameter increases, the control performance gets closer to the optimal one computed by a QP solver which is implemented by a standard package CVXOPT \cite{CVXPY_code}. This result is consistent with the suboptimal bound in Lemma \ref{lemma_suboptimality}.
It is also shown that the average computation time of prediction-correction laws (PCLs) is always smaller than the one of the QP solver. Importantly, it is noticed that the average computation time of the original QP solver is longer than the sampling step of discretization, implying that the CBF-QP method is heavily restricted by the computation efficiency of a solver when the continuous update is expected.

\section{Conclusion}
In this work, an innovative suboptimal safe control scheme in the context of the CBF method is presented using the barrier-based interior point method. Two descent-based dynamics with additional prediction terms by further exploiting the system dynamics are designed, generating cheap solutions in a continuous manner, instead of repeatedly solving the QP problems. Both algorithms are proven to converge to the suboptimal safe solutions exponentially. The key contribution of this work is the theoretical analysis of the alternative safe controller generated by prediction-correction dynamics, including suboptimality, convergence, safety, and input-to-state stability against disturbances. 

This new scheme, with the potentially better ability for online and continuous control, can be beneficial to practical applications in at least two aspects. First, since the control input is generated by closed-form dynamics, which is more suitable being embedding in low-level controllers and further accelerating on hardware processing unit \cite{YanW12, MansooriE21}. On the other hand, the myopic fixed objective functions considered in the original CBF method \eqref{constraint_ploblem_origin} can be extended to other state-dependent and time-varying objective functions \eqref{unconstrained_problem}, without loss of suboptimality and convergence guarantee. This is an important improvement for a system controlled in dynamic environments \cite{time_varying_CO_IEEE}. Therefore, it is expected to further develop the presented safe control scheme with practical applications.

In future research, the presented scheme can be improved in several ways. A significant improvement will be finding an alternative way to simplify or replace the inverse operation to reduce the computational complexity. However, theoretical analysis, such as safety and stability, is equally important when proposing an improved algorithm. More complicated settings, such as a more general CBF method \cite{exponential_CBF}, uncertainty estimation \cite{CBF_for_bayesian}, and verification on complex systems \cite{active_set_QP_robot}, are also worth studying in the future.

\begin{appendices}
\section{Detailed Formulation in Simulations}

\subsection{Derivation of Example 1}
\label{appendix_example1}
Consider a single obstacle labeled as $1$. The CBF-QP can be written into the following form
\begin{gather*}
    \min_{v \in \Real^{2}} v^\Tran v - 2 v_d^\Tran v \quad
    s.t. \quad a_1^\Tran(x) v - b_1(x) \leq 0,
\end{gather*}
where $a_1(x) = -{(x - x_c)}/{\norm{x - x_c}}$ and $b_1(x) =\alpha (\norm{x - x_c} - r) $. Then, the unconstrained cost function based on interior point barrier method is 
\begin{equation*}
    f_v = v^\Tran v - 2 v_d(x)^\Tran v - \frac{1}{c}\log (-a_1^\Tran(x) v + b_1(x)).
\end{equation*}
The derivative with respect to variable $v$ is formulated as
\begin{gather*}
    \nabla_{v} f_v = 2(v - v_d(x)) + \frac{a_1(x)}{c\left(b_1(x)-a_1^\Tran(x) v \right)}.
\end{gather*}
The Hessian matrix and partial derivatives with respect to $x$ are
\begin{gather*}
    \nabla_{vv} f_v = 2 \mathbf{I}_2 + \frac{a_1(x)a_1^\Tran(x)}{c\left(b_1(x)-a_1^\Tran(x) v\right)^2},\\
    \begin{split}
        \nabla_{vx} f_v = & 2 K_d^\Tran + \frac{\nabla_x a_1(x)}{c\left(b_1(x)-a_1^\Tran(x) v \right)} \\
    & - \frac{a_1(x) \left(\nabla_x b_1(x) - \nabla_x^\Tran a_1 (x) v \right)^\Tran }{c\left(b_1(x)-a_1^\Tran(x) v\right)^2}
    \end{split},
\end{gather*}
where
\begin{gather*}
    \nabla_x a_1(x) = - \frac{\mathbf{I}_2}{\norm{x-x_c}^2} +  \frac{\left(x - x_c\right)\left(x - x_c\right) ^ \Tran}{\norm{x-x_c}^3},\\
    \nabla_x b_1(x) = \alpha \frac{x - x_c}{\norm{x - x_c}}.
\end{gather*}
\subsection{Formulation in Example 2}
\label{appendix_example2}
The functions in the cart-pole system \eqref{cart_pole_system_dynamics} are 
\begin{gather*}
    f_v(\theta, \omega) = \frac{m_p \sin \theta \left( l \dot \theta^2 + g\cos \theta \right)}{m_c + m_p \sin^2{\theta}},\\
    f_\omega(\theta, \omega) = -\frac{m_p l \dot \theta^2\cos \theta\sin\theta + \left( m_c + m_p \right)g\sin\theta}{l\left(m_c + m_p \sin^2{\theta}\right)}, \\
    g_v(\theta, \omega) = \frac{1}{m_c + m_p \sin^2{\theta}}, \\
    g_\omega (\theta, \omega) = -\frac{1}{l\left(m_c + m_p \sin^2{\theta}\right)}.
\end{gather*}
The compact form of linear constraints in CBF-QP is given by $A^\Tran(q) u \leq B(q)$, where
\begin{gather*}
    A(q) =  \left[\begin{matrix}
    2 \theta g_\omega(\theta, \omega)& 1 & -1
    \end{matrix}\right],\\
    B(q) = \\
    \left[\begin{matrix}
    - 2\theta f_\omega(\theta, \omega) - 2\omega^2 - 2 \mu \theta\omega + \alpha \left( -2\theta\omega + \mu (r^2 - \theta^2) \right) \\
    u_{max}   \\
    -u_{min}
    \end{matrix}\right].
\end{gather*}

\section{Proofs of Main Results}
\subsection{Proof of Theorem \ref{theorem_convergance_gradient}}
To prove the theorem, two lemmas are introduced below.
\begin{lemma}\label{lemma_rank_1_condition}
    For any vector $\bar a \in \Real^m_y$, the matrix $\bar A \triangleq \bar a \bar a^\Tran$ is positive semi-definite with eigenvalues $\lambda_1(\bar A) = \bar a^\Tran \bar a \ge 0$ and $\lambda_i(\bar A) = 0$, $i =2,\dots, m_y$.
\end{lemma}
\par \emph{Proof.} Observe that $\bar A \bar a = \bar a \cdot (\bar a^\Tran \bar a) = \lambda_1(\bar A) \bar a$. Constructing $span \left(v_1,\dots,v_{m_y - 1}\right)$ as the orthogonal space of $\bar a$ with $\bar a^\Tran v_i = 0$ and $v_i^\Tran v_j = 0$, there is $A v_i = \bar a (\bar a^\Tran v_i) = 0 \cdot v_i$, $i, j = 1, \dots, m_y - 1, j \ne i$.
\begin{lemma}\label{lemma_convex_inverse_bound}
    The unconstrained problem \eqref{unconstrained_problem} is convex for all $x$ and $y$. The inverse of the Hessian $\nabla_{yy}^{-1} f_u(y, x, \vartheta)$ exists and is uniformly bounded.
\end{lemma}
\par \emph{Proof.} Notice that the constraints $F(x(t), y)$ in CBF-CLF problem \eqref{constraint_ploblem_origin} and \eqref{constraint_ploblem_QP_saturation} are linear in $y$. Splitting $A(x) = \left[ a_1(x),\dots, a_p(x)\right]$ where $a_i\in \Real^{m_y}$, $i= 1, \dots, p$, the Hessian of the logarithmic barrier function $\phi(y, x)$ with respect to $y$ is calculated by $\nabla_{yy} \phi(x,y) =  \sum_{i=1}^p \frac{1}{c f_i^2} a_i(x) a_i(x)^\Tran$. Denote $\hat a_i(x) = a_i(x) / f_i$ and $A_i(x) = \hat a_i(x) \hat a_i(x)^\Tran$, one has $\nabla_{yy} \phi(y, x) = c^{-1} \sum_{i = 1}^p A_i(x)$. Since $c>0$, according to Lemma \ref{lemma_rank_1_condition}, $\nabla_{yy} \phi (y, x)$ is positive semi-definite for all $x$ and $y$. Therefore, the Hessian of $f_u(y, x, \vartheta) = \nabla_{yy} f_0 + \nabla_{yy} \phi(y, x)$ is positive definite and $\nabla^{-1}_{yy} f_u(y, x, \vartheta)$ exists. From Assumption \ref{assump_convex_feasible}, it is obtained that $\nabla_{yy} f_u(y, x, \vartheta) \ge q_c \mathbf{I}_{m_y}$ which implies $\nabla^{-1}_{yy} f_u(y, x, \vartheta)$ is uniformly bounded for all $y$, $x$ and $\vartheta$.

\par \emph{Proof of Theorem \ref{theorem_convergance_gradient}.} The convergence of both methods can be proven by the Lyapunov method. Since problem \eqref{unconstrained_problem} is strong convex for all $y$, $x$ and $\vartheta$, the optimal solution is obtained as long as the gradient $\nabla_{y} f_u(y, x, \vartheta) = 0$. Therefore, construct the Lyapunov function as \begin{equation}
V(\sigma(t), t) =\frac{1}{2} \sigma^\Tran (t)\sigma(t), \label{Lyapunov_function}
\end{equation}
in which $\sigma(t) \triangleq \nabla_y f_u (y(t), x(t), \vartheta(x(t),t))$. Note that this abbreviation is to emphasize that $V$ is time-varying. Calculating the time derivative of $V$ yields
\begin{equation}
    \dot V(\sigma(t), t) = \sigma^\Tran(t) \left(\nabla_{y}\sigma(t) \dot y + \nabla_{x}\sigma(t)\dot x + \nabla_{\vartheta}\sigma(t) \dot \vartheta\right), \label{lyapunov_derivative}
\end{equation}
with $\nabla_t V(\sigma(t), t) = 0$. Substituting \eqref{gradient_PClaw_continuous} in \eqref{lyapunov_derivative} and following Assumption \ref{assump_convex_feasible} and Lemma \ref{lemma_convex_inverse_bound}, one has
\begin{align}
    \dot V(\sigma(t), t) =& -\gamma\sigma^\Tran(t) \nabla_{yy}f_u(y, x, \vartheta) \sigma(t)\nonumber\\
    = & -\gamma\sigma^\Tran(t)\left( \nabla_{yy}f_0(y, x, \vartheta) + \sum_{i = 1}^p \frac{A_i(x)}{c} \right)\sigma(t)\nonumber\\
    \leq & -\gamma\sigma^\Tran(t) \nabla_{yy}f_0(y, x, \vartheta) \sigma(t) \nonumber\\
    \leq& -q_c \gamma\sigma^\Tran(t) \sigma(t) = - 2q_c \gamma V(\sigma(t), t). \label{gradient_convergence_rate}
\end{align}
Solving the inequality above using comparison principle, there is $V(\sigma(t), t) \leq V(\sigma(t_0), t_0) e^{-2 q_c\gamma (t - t_0)}$. According to \cite{nonlinear_systems}, one has $\norm{\sigma (t)} \leq \norm{\sigma(t_0)} e^{- q_c\gamma (t - t_0)}$. Since $\nabla_{yy}^{-1}f_u(y, x, \vartheta)$ is uniformly bounded as proven in Lemma \ref{lemma_convex_inverse_bound}, there is
\begin{align*}
    \norm{y - y^*} = &  \norm{\nabla_{yy}^{-1}f_u(y, x, \vartheta)\nabla_{y}f_u(y, x, \vartheta)}\\
    \leq&  \frac{1}{q_c} \norm{\nabla_{y}f_u(y, x, \vartheta)}
     \leq  \frac{1}{q_c}\norm{\sigma(t_0)} e^{q_c\gamma(t - t_0)},
\end{align*}
where the first line is derived from the mean-value theorem with the fact that $\nabla_x f(x) = \nabla_{xx} f(x^\prime) (x - x^*)$ in which $x^\prime$ is the combination of $x$ and $x^*$.

\par Now, consider the Newton-based dynamics \eqref{Newton_PClaw_continuous} with the same Lyapunov function \eqref{Lyapunov_function} whose derivative becomes
\begin{align}
    \dot V(\sigma(t), t) =& -\gamma\sigma^\Tran(t) \sigma(t) = - 2\gamma V(\sigma(t), t), \label{newton_convergence_rate}
\end{align}
indicating $V(\sigma(t), t) = V(\sigma(t_0), t_0) e ^{-2 \gamma (t - t_c)}$, $\norm{\sigma (t)} \leq \norm{\sigma(t_0)} e^{- \gamma (t - t_0)}$ and $\norm{y - y^*}\leq \frac{1}{q_c}\norm{\sigma(t_0)} e^{\gamma(t - t_0)}$. The proof is complete.

\subsection{Proof of Theorem \ref{theorem_safety}}
\par\emph{Proof.} Noticing that $h_N(x) > 0$ indicates $h(x) > 0$ since $\norm{u - u^*} \ge 0$ hold all the time, it is sufficient to prove $h_N(x) > 0$ to reveal the safety of systems. In other words, it only needs to prove that $\mathcal{C}_N$ is forward invariant. Following the same process as the original CBF method, the derivative of $h_N(x)$ is
\begin{align*}
    \dot h_N(x) = & - \frac{d}{dt} \norm{\nabla_y f_u (y, x, \vartheta)} + \alpha^\prime \nabla_x h(x) \dot x\\
    \ge & \hat \gamma \norm{\nabla_y f_u (y, x, \vartheta)} \\
    & + \alpha^\prime \left(L_fh(x) + L_gh(x) \left(u^* + u - u^*\right)\right)\\
    \ge & q_c\hat\gamma  \norm{u - u^*} - \alpha^\prime \alpha h(x) + \alpha^\prime L_gh(x) (u - u^*)\\
    \ge & q_c( \hat \gamma - \alpha) \norm{u- u^*} - \alpha h_N(x) - \alpha^\prime b_h\norm{u - u^*}.
\end{align*}
Note the second line utilizes the results in Theorem \ref{theorem_convergance_gradient}. Since $\alpha^\prime =q_c (\hat \gamma - \alpha)/ b_h$, the above formula reduces to $\dot h_N(x) \ge - \alpha h_N(x)$, which implies that the set $\mathcal{C}_N$ is forward invariant according to Definition \ref{definition_CBF}. The proof is complete.

\subsection{Proof of Theorem \ref{theorem_ISS_with_disturbances}}
\par\emph{Proof.} The gradient-based dynamics is considered first. Following the same study in Theorem \ref{theorem_convergance_gradient}, the derivative of the Lyapunov function is
\begin{align}
    \dot V(\sigma(t), t) =& -\gamma\sigma^\Tran(t) \nabla_{yy}f_u(y, x, \vartheta) \sigma(t) \nonumber\\
    & +\sigma^\Tran(t) \left(\nabla_{yt}f_u(y, x, \vartheta) -  \widehat{\nabla_{yt}f_u}(y, x, \vartheta)\right) \nonumber\\
    \leq& -q_c \gamma\sigma^\Tran(t) \sigma(t) \nonumber\\
    & +\norm{\sigma(t)} \norm{\nabla_{yt}f_u(y, x, \vartheta) -  \widehat{\nabla_{yt}f_u}(y, x, \vartheta)} \nonumber\\
    \leq & -q_c \gamma (1 - \xi) \sigma^\Tran(t) \sigma(t) \nonumber\\
    & - \norm{\sigma(t)}(\xi \norm{\sigma(t)} - \sup\norm{\varrho(t)}). 
    \label{lyapunov_derivative_temp}
\end{align}
When $\xi \norm{\sigma(t)} \ge \sup \norm{\varrho(t)}$. Then, one has
\begin{equation*}
    \dot V(\sigma(t), t) \leq - 2 q_c \gamma (1 - \xi) V(\sigma(t), t),
\end{equation*}
implying that $\norm{\sigma(t)}$ is ultimately bounded by $\sup\norm{\varrho(t)}/\xi$. Therefore, by $\norm{y - y^*} \leq \frac{1}{q_c} \norm{\sigma(t)}$, the approximate error is ultimately bounded by $\sup\norm{\varrho(t)}/(\xi q_c)$. It is easy for extension to prove the ISS property and ultimate bound of the systems \eqref{Newton_PClaw_continuous}. The repeated proof is omitted here.

\subsection{Proof of Corollary \ref{colollary_safe_with_disturbances}}
\par \emph{Proof.} Using the results in Theorem \ref{theorem_ISS_with_disturbances} together with \eqref{quantitative_describe_evolution} and \eqref{disturbances_cbf}, one has
\begin{align*}
    \dot {\widetilde h}_N(x) \ge & q_c \hat \gamma \norm{u - u^*} - \sup \norm{\varrho(t)}\\
    &+ \alpha^\prime \left(L_f \widetilde h(x) + L_g \widetilde h(x) \left(u^* + u - u^*\right) + L_d \widetilde h(x)\right)\\
    \ge & q_c \hat \gamma \norm{u - u^*} - \sup \norm{\varrho(t)}
    + \alpha^\prime \alpha\widetilde h(x)\\
    & + \alpha^\prime L_g \widetilde h(x)(u - u^*)\\
    \ge & q_c( \hat \gamma - \alpha) \norm{u- u^*} - \alpha h_N(x) - \alpha^\prime b_h\norm{u - u^*}\\
    \ge & \alpha h_N(x).
\end{align*}
Therefore, the set $\widetilde{\mathcal{C}}_N $ is forward invariant, such that the safety of systems is ensured as long as $(x(t_0), y(t_0)) \in \widetilde{\mathcal{C}}_N$.
\end{appendices}

{
\bibliographystyle{IEEEtran}
\bibliography{mybib}
}

\end{document}